\documentclass[aps,prl,twocolumn,showpacs,nobibnotes,reprint]{revtex4-1}
\pdfoutput=1
\usepackage{graphics,graphicx,amsfonts,amsmath,amsbsy,amssymb,color}
\usepackage{verbatim}  
\usepackage{psfrag}  
\usepackage{tabularx}
\usepackage{xmpmulti}
\usepackage{hyperref}
\usepackage{marginnote}
\usepackage{subfigure}
\usepackage{paracol}
\usepackage{xparse}
\usepackage{layouts}
\usepackage{afterpage}
\usepackage{braket}
\usepackage{paralist}
\usepackage{bm}
\usepackage{url}
\newcommand{\refeq}[1]{{Eq.~(\ref{#1})}}
\newcommand{\reffig}[1]{{Fig.~\ref{#1}}}

\newcommand{\Ndet}{\ensuremath{N_{\text{det}}}}
\newcommand{\Nelec}{\ensuremath{N_{\text{elec}}}}
\newcommand{\Norb}{\ensuremath{N_{\text{orb}}}}
\newcommand{\Ncrit}{\ensuremath{N_\text{plat}}}

\hypersetup{hyperindex,breaklinks,colorlinks=true,linkcolor=blue,citecolor=blue,urlcolor=blue,filecolor=blue}

\begin{document}

\title{The sign problem in full configuration interaction quantum Monte Carlo: \\ Linear and sub-linear representation regimes for the exact wave function}

\author{James~J.~Shepherd$^{(a)}$}
\author{Gustavo~E.~Scuseria$^{(a)}$}
\author{James~S.~Spencer$^{(b)}$}

\address{$^{(a)}$ Department of Chemistry and Department of Physics and Astronomy, Rice University, Houston, TX 77005-1892 \\ $^{(b)}$ Department of Materials, Imperial College London, Exhibition Road, London, SW7 2AZ, U.K. and Department of Physics, Imperial College London, Exhibition Road, London, SW7 2AZ, U.K.}

\begin{abstract}
We investigate the sign problem for full configuration interaction quantum Monte Carlo (FCIQMC), a stochastic algorithm for finding the ground state solution of the Schr\"odinger equation with substantially reduced computational cost compared with exact diagonalisation. 
We find $k$-space Hubbard models for which the solution is yielded with storage that grows sub-linearly in the size of the many-body Hilbert space, in spite of using a wave function that is simply linear combination of states.
The FCIQMC algorithm is able to find this sub-linear scaling regime without bias and with only a choice of Hamiltonian basis.
By means of a demonstration we solve for the energy of a 70-site half-filled system (with a space of $10^{38}$ determinants) in 250 core hours, substantially quicker than the $\sim$10$^{36}$ core hours that would be required by exact diagonalisation. This is the largest space that has been sampled in an unbiased fashion.
The challenge for the recently-developed FCIQMC method is made clear: expand the sub-linear scaling regime whilst retaining exact on average accuracy.
This result rationalizes the success of the initiator adaptation (i-FCIQMC) and offers clues to improve it. 
We argue that our results changes the landscape for development of FCIQMC and related methods.
\end{abstract}
\date{\today}
\maketitle

\afterpage{\afterpage{%
\onecolumngrid%

\begin{figure}
\vspace{-0.5cm}
\hspace*{-0.05\textwidth}\makebox[1.2\textwidth][l]{%
\subfigure[\mbox{}]{%
\includegraphics{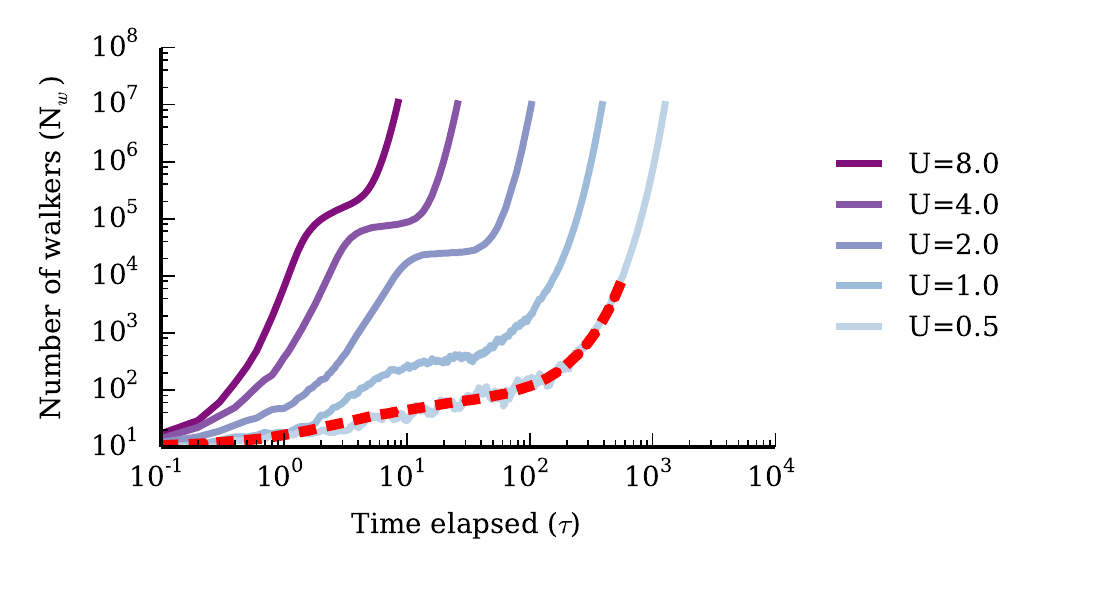}\label{one}
}\hspace*{-0.03\textwidth}%
\subfigure[\mbox{}]{%
\includegraphics{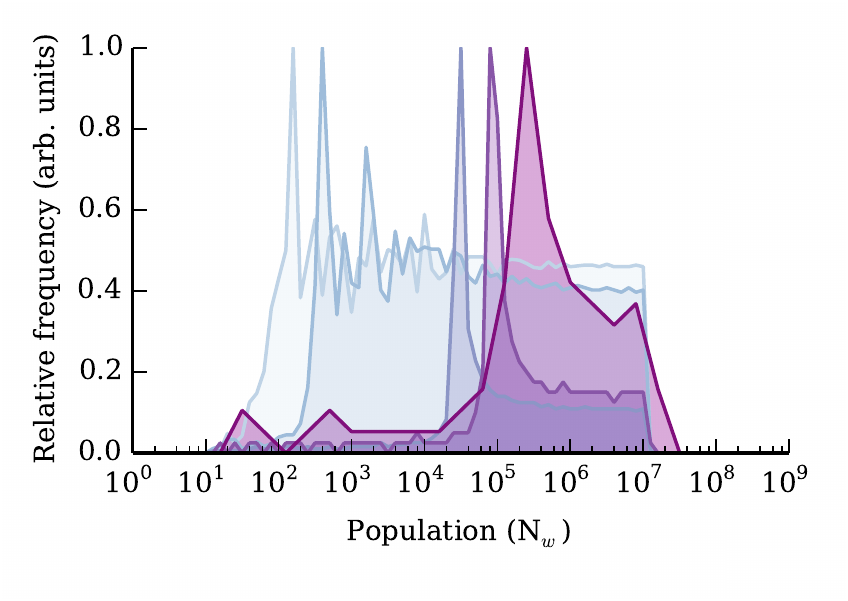}\label{two}
}
}
\caption{(Colour online). These two panels describe how we are able to identify plateaus. In (a), the decrease in $U$ is shown to obscure the plateau. The bold, black (red online), dashed annotation indicates the $U=0.5$ population growth averaged over 100 random number seeds.
In (b), we show that the population at the plateau corresponds to the maximum value of histogram of the population (although, in general, variable bin widths were used).  These two panels share a common key.}
\end{figure}

\twocolumngrid
}}

\afterpage{\afterpage{\afterpage{\afterpage{%
\onecolumngrid %

\begin{figure}
\vspace{-1.0cm}
\hspace*{-0.05\textwidth}\makebox[1.2\textwidth][l]{%
\subfigure[\mbox{}]{%
\includegraphics{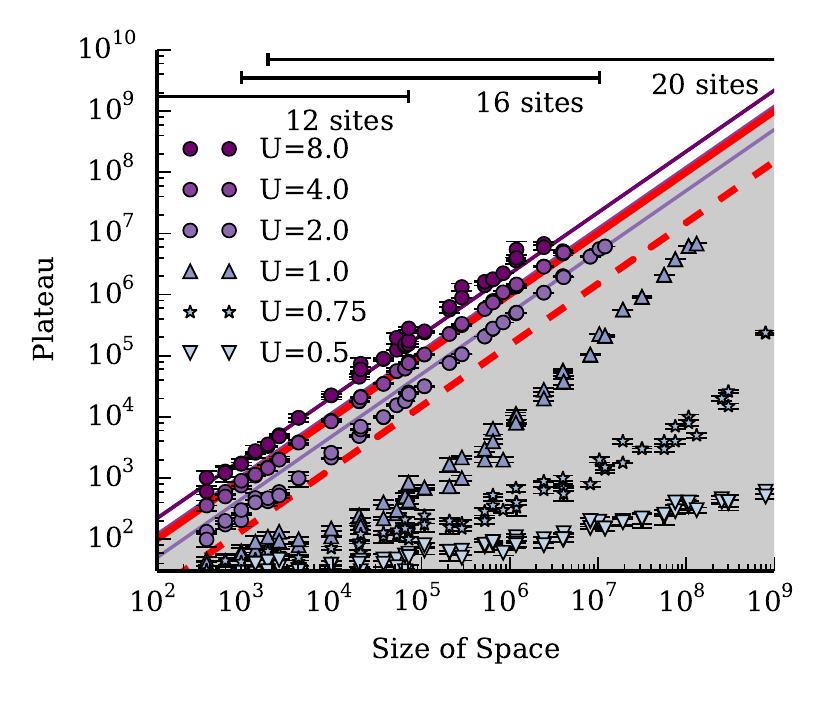}\label{three}
}\hspace*{-0.03\textwidth}
\subfigure[\mbox{}]{ 
\includegraphics{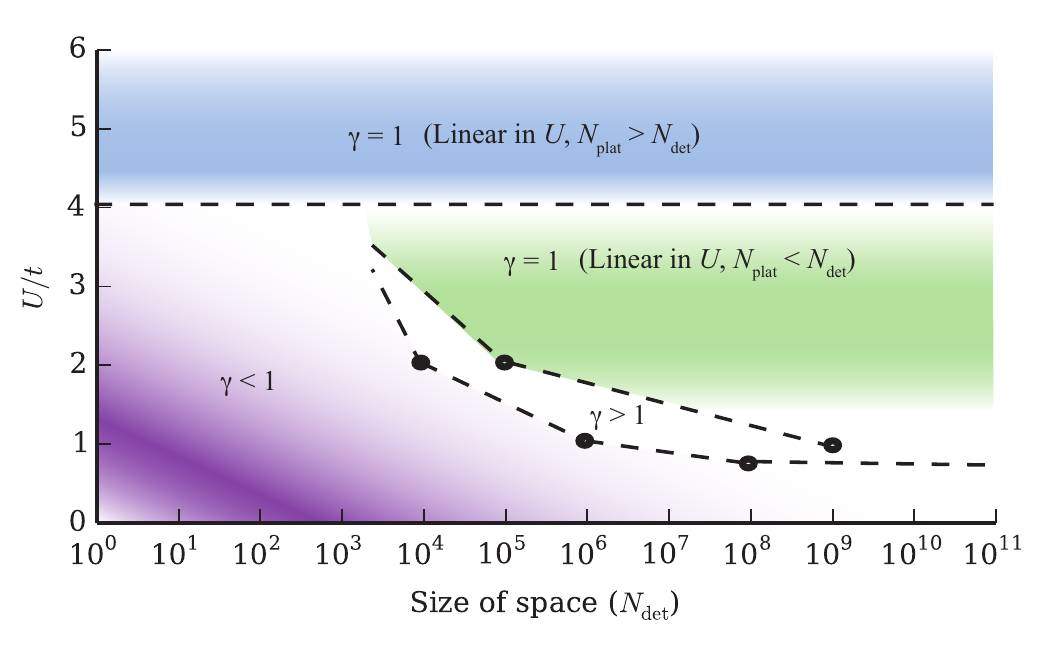}\label{four}}%
}
\caption{(Colour online). Analysis of all plateaus found in this study. In (a), plateau heights are plotted against sizes of space. In (b) the scaling trends are summarised in a diagram relating it to system parameters. 
In (a), the following bold, black (red online) annotations have been made. 
The solid line indicates the $\Ncrit=\Ndet$, where we are storing one walker per determinant (on average) across the simulation and the storage becomes comparable to FCI (unshaded region). 
The dashed line indicates where $U=1.0$ would be if it retained the same plateau height scaling as high $U$ values. 
The size of space was calculated by a Monte Carlo method and takes into account momentum symmetry~\cite{booth_thesis}.
}
\end{figure}

\twocolumngrid
}}}}

{\bf \emph{Introduction.--} }
Exact methods for solving the Schrodinger equation are used at the forefront of understanding in condensed matter physics~\cite{capriotti_spin-liquid_2004,daghofer_model_2008,maier_dynamics_2008,yan_spin-liquid_2011} and in molecular quantum chemistry~\cite{booth_towards_2013,kurashige_entangled_2013,sharma_low-energy_2014}.
However, exact parameterizations of the many-body wave function for a general system of interacting fermions scale exponentially with the system size, \emph{i.e.} $\mathcal{O}(  e^N )$.
Quantum Monte Carlo (QMC) techniques attempting to determine these parameters are hindered by their values being either positive or negative, causing more pronounced variability than a set of parameters with a single sign. 
Unconstrained algorithms~\cite{ceperley_ground_1980} therefore also scale exponentially. Constrained algorithms, in which the sign of the wavefunction parameters are fixed (\emph{e.g.}\ by a trial wavefunction), can be polynomially-scaling but at the cost of a bias~\cite{zhang_exact_1991,zhang_constrained_1995,foulkes_quantum_2001,capello_superfluid_2007}.
The (unsolved) challenge to find a constraint on the signs of a wave function, whilst still reproducing the exact result, is usually termed the fermion sign problem. 
For some QMC methods and specific systems this constraint can be imposed exactly, abstracting away the fundamental complexity of the problem.
Overcoming the sign problem is vital for the accurate treatment of real systems.

Here, we investigate the sign problem of a recently-developed QMC method developed for use in finite molecular basis sets: full configuration interaction QMC (FCIQMC)~\cite{booth_towards_2013}. 
This is the direct (ground-state) QMC analogue of exact diagonalisation, finding the exact lowest-energy solution for a finite Hilbert space with an exponential number of states using a walker-based algorithm, where the Hilbert space is a set of Slater determinants and grows exponentially with the number of fermions.
Since it does not impose the signs of the wave function in advance, this method does in general have a sign problem~\cite{spencer_sign_2012}, and the cost of the storage of the exact-on-average wave function has been shown to scale linearly in the size of the Hilbert space for a series of atomic systems~\cite{booth_fermion_2009,booth_approaching_2010,cleland_study_2011}.
Inspired by recent interest in high-throughput data driven informatics~\cite{jain_high-throughput_2011} we here study 378 systems with a plateau using a high-throughput approach. 
In so doing, we extend the information available about this method considerably.

As shown below, we discover a regime of the $k$-space 1D Hubbard model where the amount of information required to store the ground-state wave vector has sub-linear scaling with size of the Hilbert space states.
This is achieved during the simulation in the presence of a sign problem, assuming a linear wave function ansatz, without requiring any information or bias beyond the Hamiltonian.
This regime is smoothly connected to the more typical linear-scaling regime including situations in which there are more walkers required than the number of states. 
We use this to build a conceptual map based on the system's parameter space (Hubbard $U$ and size of Hilbert space) for the regions of scaling for this promising QMC technique. 
We relate these findings back to FCIQMC and its initiator adaptation, providing concrete insight for the development of these methods.
We discuss whether, in light of this, QMC for the FCI problem could receive routine use for the treatment of correlated electrons in many more realistic contexts. 

{\bf \emph{Theory.--}} FCIQMC, like all projector-based methods, exploits the fact that $\ket{\Psi(\tau)} = e^{-\tau\hat{H}} \ket{\Psi(0)}$ tends to the ground-state solution of the imaginary time Schro\"dinger equation in the limit of $\tau\rightarrow\infty$ if $\ket{\Psi(0)}$ has a non-zero overlap with the ground-state.  The configuration interaction wavefunction ansatz $\ket{\Psi} = \sum_{\bm i} c_{\bm{i}} \ket{D_{\bm{i}}}$ is used, where $\{\ket{D_{\bm{i}}}\}$ is a set of Slater determinants of size \Ndet{} formed from \Nelec{} electrons and \Norb{} spin-orbitals.  A first-order Euler finite-difference approximation to the imaginary-time Schro\"dinger equation gives
\begin{equation}
    c_{\bm{i}}(\tau+\delta\tau) = c_{\bm{i}}(\tau) - \delta\tau \sum_{\bm{j}} H_{\bm{ij}} c_{\bm{j}}(\tau),
        \label{eqn:fciqmc}
\end{equation}
where $H_{\bm{ij}} = \braket{D_{\bm{i}}|\hat{H}-S|D_{\bm{j}}}$ and an energy offset (`shift'), $S$, has been introduced in order to conserve normalisation.  For a sufficiently small timestep~\cite{spencer_sign_2012}, $\delta\tau$, the coefficients tend to the ground-state of the Hamiltonian matrix. Although FCIQMC is essentially a stochastic version of the power method~\cite{spencer_sign_2012,petruzielo_semistochastic_2012}, the core algorithm has inspired a wide range of developmental advances and new methods~\footnote{These include CCMC~\cite{thom_stochastic_2010}, SQMC~\cite{petruzielo_semistochastic_2012}, DMQMC~\cite{blunt_density-matrix_2014}, FCIQMC for excited states~\cite{booth_communication:_2012} and MSQMC~\cite{ten-no_stochastic_2013}}.

The wavefunction coefficients are discretized by representing them with a set of signed walkers~\footnote{It has recently been shown that using walkers with non-integer weights can be more efficient~\cite{petruzielo_semistochastic_2012}}.
In each timestep, the set of walkers is considered each in turn and \refeq{eqn:fciqmc} sampled according to unbiased rules~\cite{booth_fermion_2009}. 
Since the simulation allows the sign of a site to change, because the off-diagonal matrix elements are of different signs, 
the arithmetic that occurs on a site to determine its overall sign involves a process termed annihilation, removing pairs of walkers which belong to the same determinant but have opposite signs.
The annihilation step preserves the expected distribution of walkers and crucially prevents the growth of exponential noise and collapse onto the sign-problem-free ground state of the matrix defined by $\tilde{H}_{\bm{ij}} = H_{\bm{ij}} \delta_{\bm{ij}} - (1-\delta_{\bm{ij}}) \left|H_{\bm{ij}}\right|$~\cite{spencer_sign_2012}.

The sign problem arises in FCIQMC because the signs of the FCI coefficients are not known in advance. 
If they were known, it is postulated that the sign problem could be removed by factorization~\cite{kolodrubetz_partial_2012,Roggero,Mukherjee}. 
In contrast the matrix $\tilde{H}$ has coefficients that are all of the same sign in much the same way as a bosonic wavefunction has the same sign in its value for all particle coordinates. 
This is the determinant space analogue of the real-space bosonic solution for FCIQMC~\cite{kolodrubetz_effect_2013}. 

A typical simulation contains four distinct phases~\footnote{Here, we have concentrated on features which are salient to our investigation; for full details refer to (\emph{e.g.}) Refs.~\cite{booth_fermion_2009,cleland_communications:_2010,booth_breaking_2011,shepherd_investigation_2012,spencer_sign_2012,petruzielo_semistochastic_2012,booth_towards_2013,booth_linear-scaling_2014}}.
Initially the shift is held constant (typically to a mean-field energy) and the population of walkers grows exponentially.  The population spontaneously stops growing and enters the \emph{plateau} phase at a system-dependent population, during which the ground-state sign structure emerges. The population spontaneously begins to grow again at an exponential rate, albeit slower than before.  
The shift is then varied to keep the population approximately constant. 
Above the plateau, statistics can be accumulated that are demonstably from the exact solution\cite{booth_fermion_2009}; the post-plateau population is a stochastic representation of the exact wave function. 
The first three phases can be seen in \reffig{one}.

{\bf \emph{Plateau determination.--} } The plateau is therefore a very powerful conceptual feature of FCIQMC\@. 
Phenomenologically, the plateau provides an unambiguous signal of how hard the sign problem is because it represents the minimum storage cost for an on-average exact representation of the FCI vector~\cite{booth_fermion_2009,spencer_sign_2012}. 
Computationally, this number of walkers determines the dominant scaling bottleneck, for both memory and computer time, of the method since each Monte Carlo iteration loops over this list. 
The stochastic sampling of the propagator will also contribute to the noise of the simulation %
but this is pre-multiplied by the length of the main vector. 

Crucially therefore (and uniquely in projector QMC methods) the plateau provides an unambiguous measure of the sign problem in FCIQMC: by comparing the plateau height against the number of determinants in the Hilbert space, we obtain a measure for how `hard' a system is for FCIQMC.

In order to study plateau heights, it is important to establish a unique and reproducible definition accounting for variations seen in \reffig{one}, where plateaus are obscured by becoming `shoulder'-shaped or being overwhelmed by stochastic noise.
The plateau can be thought of as the walker population that the simulation spends the most time at. 
To find this, the relative frequency that a certain population window appears in the simulation is computed and the maximum of this distribution found.
The histogram of the logarithm of the population rather than that of the population is used in order to handle the exponential growth in population.
The plateau signal is shown for various values of $U$ in \reffig{two}. 
The disadvantage of this approach is that for some values of $U$, this can lead to \emph{overestimation} of the plateau for some runs as multiple peaks compete.
This can be circumvented by changing the bin width, and this must sometimes be interpreted manually. 
This procedure is discussed in more detail in the Supplementary Material, where each plateau can also be verified by inspection~\footnote{See Supplemental Material at [URL will be inserted by publisher], which contains finer details on the histogram method for determining plateau height and population graphs
for every system consider and includes Refs.~\onlinecite{shepherd_emergence_2012,figshare_repository1,*figshare_repository2,figshare_repository_jss}.}.

{\bf \emph{Plateau analysis.--} }  We now consider the 1D translationally invariant ($k$-point) one-band Hubbard model, for the parameter range $U=$ 0.5, 0.75, 1.0, 2.0, 4.0, and 8.0 for $N_s=$ 12, 14, 16, 18, 20, and 22 sites per simulation cell. We explore a wide range of doping levels (4 to $2 N_s-4$) for consistent simulation parameters~\footnote{In general we fix the initial shift to the correlation energy zero, the time step to be $\delta\tau=0.001$a.u and start with an initial population of 10 positive walkers on the restricted Hartree--Fock determinant. The plateaus are somewhat sensitive to the initial population and $\delta\tau$, but our general conclusions are not}. All calculations were performed with the HANDE QMC code~\cite{hande}.  We focus on 1D systems because shell-filling effects were anticipated to make interpretation substantially more difficult in higher dimensions. Although this system is only of one dimension, the range of parameters encompasses a wide range of correlation regimes. 

All of the plateaus we have found are plotted in \reffig{three}. We can use this to probe the different scaling laws, based on $\Ncrit = \beta \Ndet^{\gamma}$,
where the exponent $\gamma$ is defined by the tangent to the curve at a given $\Ndet$.
Some of the trends described below are slight and to aid readers a larger version of the graph can be found in Supplemental Material.

{\bf \emph{Linear in $\Ndet$ and in $U$. --}} This is the conventional scaling regime that has been previously observed. 
Three diagonal-running parallel lines ($\Ndet \propto \Ncrit$) fit the data from $U=$ 2.0, 4.0, and 8.0 at high $\Ndet$. 
The behaviour of the gradient with $U$ is consistent with the plateau being linear in $U$ (as shown in Ref.~\onlinecite{spencer_sign_2012}). 
Overall, therefore, $\gamma=1$ and $\beta \propto U$.
We note in passing that these trends are remarkably consistent as doping and the particle number are changed.

The bold red line, almost coincident with most of the $U=4.0$ data set, shows the line of $\Ndet=\Ncrit$ where on average we store the same number of integers as the size of the space. 
The grey shading indicates where we would expect, therefore, to store less information than the full wave vector in order to obtain the solution via an exact diagonalisation (or FCI); above this line in the unshaded region the memory requirement is comparable to FCI\@.
Although this is true for storage, the computational time is still expected to be linear in the size of the space, and this (being the upper limit of the scaling here) is better than many diagonalisation routines.

{\bf \emph{Sub-linear in $\Ndet$, non-linear in $U$. --}} At sufficiently small system sizes, sub-linear scaling ($\gamma < 1$) is observed for all $U$ except $U=8.0$. 
The region of this reduced scaling depends on $U$, and extends to larger system sizes for smaller $U$. 
The lowest measurable exponent observed is $\gamma\approx0.1$ for $U=0.5$.%
The reduced exponent is surprising for two reasons.
The first is that the FCI wave function is apparently representable with storage that is sub-linear with the size of the space.
The second is that the projector algorithm here is able to find this minimal representation with no additional information than the Hamiltonian, and in particular no biasing.

{\bf \emph{Non-linear in $\Ndet$ and in $U$. --}} In the intermediate region between these two regimes, there is a polynomial region in $\Ndet$ ($\gamma < 1$) as the scaling law seems to return to the original linear scaling regime (with no shift) and crosses through $\gamma = 1$. 
This is most prominently seen by careful examination of $U=2.0$, which only deviates from the conventional scaling laws slightly. 
The return to $\gamma = 1$ appears around $\Ndet=10^5$. 
The dashed, red line shows where we would expect the $U=1.0$ scaling to be if the linear scaling with $\Ndet$ and $U$ continued, which our data set never reaches.
Nonetheless, the limiting scaling at high $\Ndet$ seems to be linear in $\Ndet$ and $U$.

{\bf \emph{Sign problem diagram.--} } These scaling relationships are summarized with respect to the system parameters \Ndet\ and $U$ in \reffig{four}
The tie-lines we have plotted are made by hand, and are estimates limited by the breadth of our data set. 
In particular, sharp lines should be considered as estimates and not definitive.
To the best knowledge of the authors, this provides the most comprehensive summary of known information about the sign problem, and scaling, in FCIQMC.

We also observe that as the system size is raised, the method returns to linear scaling in the size of the space and exponential scaling in particle number.
This is interesting because it seems like it is a reverse to what might be expected to happen. 
As the system gets closer to the thermodynamic limit, quantities such as the energy become extensive (\emph{i.e.}\ scale linearly with $\Nelec$). As the correlation length is increasingly well-contained within the simulation cell, we would expect the problem to become easier and of improved scaling (due to self-averaging).

{\bf \emph{Conclusion \& Discussions.--} } Our principal conclusion is the discovery of a regime of the $k$-space Hubbard model where the exact ground state can be stored with sub-linear representation cost. 
This exact-on-average representation requires no prior knowledge beyond on-the-fly access to the Hamiltonian matrix elements, and we believe this finding to be widely significant~\footnote{We disclaim in passing that we have not considered stochastic error, deliberately. For a discussion of stochastic error in the polynomial-scaling diffusion Monte Carlo, see Ref.~\onlinecite{nemec_diffusion_2010}.}.
By means of a practical demonstration of the significance of this regime, we can find the ground-state energy for the half-filled 70-site system for $U=0.1$ in 250 core hours ($E=-87.418564(7)$~a.u.).
The size of space is $10^{38}$ determinants, and this is the largest unbiased simulation to date.
By way of comparison, we estimate exact diagonalisation would take 10$^{36}$ core hours, based on known scaling laws and calculations from smaller system sizes using the algorithm implemented in ALPS~\cite{albuquerque2007alps,bauer2011alps}.
This poses the question: how many more, larger, systems are available for study that have simply not yet been attempted? 

The low-scaling regime, occurring in a greater range of systems at low $U$, seems co-incident with the weakly interacting, or weakly correlated, regime. 
Although this is a tempting conclusion to draw, this is not a link that we have the scope to explore in detail here. 
This is in part due to the sign problem being representation-dependent.%
In particular, the 1D Hubbard model is a toy model, not only because it is already solvable at polynomial cost~\cite{lieb_absence_1968,white_density_1992}, but also because a transformation to the real space basis set leaves it sign problem free for FCIQMC~\cite{spencer_sign_2012}.
Nevertheless, where there exist large expanses of the Slater determinant space that are redundant, FCIQMC should be able to find them, but that sparsity must exist to be found.

Finding such representations is greatly facilitated by our study here.
This is first and foremost because we demonstrate the potential benefits to be found, but also for the resources this study provisions for development of FCIQMC.
We start a database of plateaus, semi-automated plateau height analysis and a practical understanding as to what might happen to the plateau or sign problem with further development.
We hope that these concerns are placed at the forefront of FCIQMC development.
One such route of promise and significance is the discussion of symmetry breaking and restoration in the context of QMC techniques~\cite{thomas_symmetry_2014,shi_symmetry-projected_2014,shi_symmetry_2013,jimenez-hoyos_projected_2012}.
Another is the adaptation of the core algorithm to other Fock space QMC methods~\cite{petruzielo_semistochastic_2012,ten-no_stochastic_2013,thom_stochastic_2010,blunt_density-matrix_2014}, but in the wider context of other quantum chemical methods it is important to know whether there is a sign problem at all~\cite{willow_stochastic_2014,neuhauser_expeditious_2013,willow_stochastic_2013,willow_stochastic_2012,thom_stochastic_2010}.

The appearance of a sub-linear regime is interesting because it mirrors some evidence that the initiator adaptation has this scaling in its wave function representation~\cite{cleland_communications:_2010,cleland_study_2011}. 
The initiator adaptation (i-FCIQMC) imposes a population dependence on $H_{\bm{ij}}$, zeroing some elements that are considered outside the currently well-sampled space. This greatly enlarges the size of systems that can be sampled, up to 10$^{108}$ determinants to date~\cite{shepherd_full_2012}, but at the cost of a systematically improvable bias. 
In this context, therefore, i-FCIQMC is an approximate (but systematically improvable) method that \emph{expands} the sub-linear scaling rather than this scaling being unique to this approximation. 
This strongly implies there are further improvements that can be made to expand this reduced scaling still further. 

To the wider community in QMC methods, we hope this shows that FCIQMC provides interesting phenomenology and therefore something else to offer beyond FCI-quality energies.
The analysis we present here argues that a sign problem that is easy to detect is potentially more informative than an error that is unquantifiable. 
It also demonstrates that FCIQMC does have the \emph{potential} to solve large systems, which is surely required for its application in condensed matter physics, provided that its sign problem can be controlled. 
This puts emphasis back onto understanding and solving the sign problem, which is also a more universal effort. %

{\bf \emph{Acknowledgements.--} } We are grateful to Alex Thom for comments on the manuscript and early access to Ref.~\onlinecite{thom_iccmc_2014}. 
For discussions, we thank Ali Alavi, George Booth, Matthew Foulkes and Carlos Jim\'enez-Hoyos.
JSS acknowledges the support of the Thomas Young Centre under grant TYC-101 and Imperial College High Performance Computing Service~\cite{ICHPC}.  %
JJS acknowledges Royal Commission for the Exhibition of 1851 for a Research Fellowship.
JJS and GES acknowledge grants from the Welch Foundation (C-0036), DOE-CMCSN (DE-SC0006650) and local computing resources supported in part by the Cyberinfrastructure for Computational Research funded by NSF under Grant CNS-0821727.

\end{document}